\documentclass[structabstract]{aa}  
\usepackage{graphicx}
\usepackage{txfonts}
\def\XMM{XMM-{\sl Newton}}
\def\ASCA{{\sl ASCA}}

\def\Chan{{\sl Chandra}}
\def\OVIII{O\,{\sc viii}}
\def\OVII{O\,{\sc vii}}
\def\OVI{O\,{\sc vi}}
\def\OV{O\,{\sc v}}
\def\OI{O\,{\sc i}}
\def\CVI{C\,{\sc vi}}
\def\CV{C\,{\sc v}}
\def\NVII{N\,{\sc vii}}
\def\NVI{N\,{\sc vi}}
\def\NeIX{Ne\,{\sc ix}}
\def\NeX{Ne\,{\sc x}}
\def\SiXI{Si\,{\sc xi}}
\def\FeXVII{Fe\,{\sc xvii}}
\def\FeXVIII{Fe\,{\sc xviii}}

\begin{document}
   \title{\XMM{} RGS observation of the warm absorber in Mrk 279}


   \author{J. Ebrero\inst{1}
          \and
          E. Costantini\inst{1}
          \and
          J. S. Kaastra\inst{1,2}
          \and
          R. G. Detmers\inst{1}
          \and
          N. Arav\inst{3}
          \and
          G. A. Kriss\inst{4}
          \and
          K. T. Korista\inst{5}
          \and
          K. C. Steenbrugge\inst{6,7}
          }

   \institute{SRON Netherlands Institute for Space Research, Sorbonnelaan 2, 3584 CA, Utrecht, The Netherlands\\
              \email{J.Ebrero.Carrero@sron.nl}
         \and
             Astronomical Institute, University of Utrecht, Postbus 80000, 3508 TA, Utrecht, The Netherlands
             \and
             Department of Physics, Virginia Tech, Blacksburg, VA 24061, USA
             \and
             Space Telescope Science Institute, 3700 San Martin Drive, Baltimore, MD 21218, USA
             \and
             Department of Physics, Western Michigan University, Kalamazoo, MI 49008, USA
             \and
             Instituto de Astronom\'ia, Universidad Cat\'olica del Norte, Avenida Angamos 0610, Casilla 1280, Antofagasta, Chile
             \and
             University of Oxford, Department of Physics, Keble Road, OX1 3RH, Oxford, UK
             }

   \date{Received <date>; accepted <date>}

 
  \abstract
   {The Seyfert 1 galaxy Mrk 279 was observed by \XMM{} in November 2005 in three consecutive orbits, showing significant short-scale variability (average soft band variation in flux $\sim$20\%). The source is known to host a two-component warm absorber with distinct ionisation states from a previous \Chan{} observation.}
   {To study the warm absorber in Mrk 279 and investigate any possible response to the short-term variations of the ionising flux, and to assess whether it has varied on a long-term timescale with respect to the \Chan{} observation.}
   {The \XMM{}-RGS spectra of Mrk 279 were analysed in both the high- and low-flux states using the {{\tt SPEX}} fitting package.}
   {We find no significant changes in the warm absorber on neither short timescales ($\sim$2 days) nor at longer timescales (two and a half years), as the variations in the ionic column densities of the most relevant elements are below the 90\% confidence level. The variations could still be present but are statistically undetected given the signal-to-noise ratio of the data. Starting from reasonable standard assumptions we estimate the location of the absorbing gas, which is likely to be associated with the putative dusty torus rather than with the Broad Line Region if the outflowing gas is moving at the escape velocity or larger.}
   {}


   \authorrunning{J. Ebrero et al.}
   \titlerunning{The warm absorber of Mrk 279}

   \maketitle
%

\section{Introduction}
\label{intro}

It is widely believed that active galactic nuclei (AGN) are powered by gravitational accretion of matter onto the supermassive black hole that resides in their centres (see e.g. Rees \cite{Rees84} for a review). X-ray emission, characteristic of AGN activity, is therefore a fundamental tracer of the accretion processes that take place in the innermost regions of the AGN engine.

More than 50\% of the Seyfert 1 galaxies exhibit clear signatures of a photoionised gas, the so-called warm absorber (Reynolds \cite{Rey97}, George et al. \cite{Geo98}), in their X-ray spectra in the form of narrow absorption features, usually blueshifted by a few hundreds of km s$^{-1}$, thus revealing the presence of outflows from the nucleus along the line of sight (Crenshaw et al. \cite{Cren99}, Kaastra et al. \cite{Kaa00}, Kaspi et al. \cite{Kas01}, Kaastra et al. \cite{Kaa02}). The study of warm absorbers (WA, hereafter) provides a unique insight to the innermost AGN environment, allowing us to probe e.g. elemental abundances, degrees of ionisation, outflow velocities.

However, the geometrical structure and origin of the WA as well as its physical connection with the continuum emission source and other components of the AGN such as the broad line region (BLR), narrow line region (NLR) and the putative dusty torus, is not yet clear. A number of theories have been proposed to explain the origin of WA such as evaporation from the torus (Krolik \& Kriss \cite{KK01}), outflowing winds from the accretion disc (e.g. Murray \& Chiang \cite{MC95}, Elvis \cite{Elvis00}), or the ionisation cone theory (Kinkhabwala et al. \cite{Kink02}). The study of Blustin et al. (\cite{Blustin05}) on 23 AGN suggests that the WA in most of the nearby Seyfert galaxies is likely to be originated in outflows from the molecular torus.

The main problem to determine the distance between the WA and the ionising source comes from the fact that the determination of the ionisation parameter of the gas and the ionising luminosity from a single time-averaged observation only allows to derive the product $n_eR^2$. This degeneracy of the electron density $n_e$ and the distance $R$ can be broken by measuring the variations in the ionisation state of the gas in response to changes in the ionising continuum. Therefore, the density of the gas can be measured and its distance from the source determined (e.g. NGC 3516, Netzer et al. \cite{Net02}). Studies of variability of the WA have been carried out only in a handful of objects on time scales of the order of a few ks (e.g. MCG-6-30-15, Gibson et al. \cite{Gib07}, Miller et al. \cite{Miller08}; NGC 4051, Krongold et al. \cite{Kron07}; NGC 1365, Risaliti et al. \cite{Ris09}) and also on longer timescales (e.g. NGC 5548, Krongold et al. \cite{Kron10}).

Mrk 279 is a nearby ($z=0.0305$, Scott et al. \cite{Scott04}) Seyfert 1 galaxy ($F_{0.5-2\,{\rm{~keV}}}\sim 2\times 10^{-11}$~erg cm$^{-2}$~s$^{-1}$, this work). The source was previously observed in X-rays by the {\sl HEAO 1 A-2} experiment (Weaver et al. \cite{Wea95}) and \ASCA{} (Weaver et al. \cite{Wea01}) in order to study the spectral complexity of the Fe K$\alpha$ emission line region. More recently, Mrk 279 was observed in X-rays using \Chan{}-HETGS simultaneously with {\sl FUSE} and {\sl HST-STIS} in 2002 (Scott et al. \cite{Scott04}). They found strong complex absorption features from both high- and low-ionisation elements in the UV spectrum but no hints of absorption in the X-ray band due to the low continuum flux at the moment of the observation. In 2003 the source was observed again by {\sl FUSE}, {\sl HST-STIS} (Gabel et al. \cite{Gabel05}), and LETGS onboard \Chan{} (Costantini et al. \cite{cos07}). They quantified the contribution of the BLR to the soft X-ray spectrum, via the locally optimally emitting clouds (LOC) model (Baldwin et al. \cite{Bal95}), and determined that the broad lines observed at soft X-rays were consistent with a BLR origin.

This paper is devoted to the \XMM{}-RGS spectra of Mrk 279. The \XMM{}-EPIC broad band analysis is presented in a companion paper (Costantini et al. \cite{cos09}, Paper I hereafter).

This paper is organised as follows. In Sect.~\ref{data} we describe the X-ray datasets used in this study. In Sect.~\ref{analysis} we describe the spectral analysis of the most relevant features, while in Sect.~\ref{discussion} we discuss our results. Finally, our conclusions are reported in Sect.~\ref{conclusions}.

Throughout this paper we have assumed a cosmological framework with $H_0=70$~km s$^{-1}$ Mpc$^{-1}$, $\Omega_M=0.3$ and $\Omega_\Lambda=0.7$ (Spergel at al. \cite{Spergel03}). The quoted errors refer to 68.3\% confidence level ($\Delta C = 1$ for one parameter of interest) unless otherwise stated.


\section{The X-ray data}
\label{data}

\begin{figure}
\centering
\includegraphics[width=6.5cm,angle=90.]{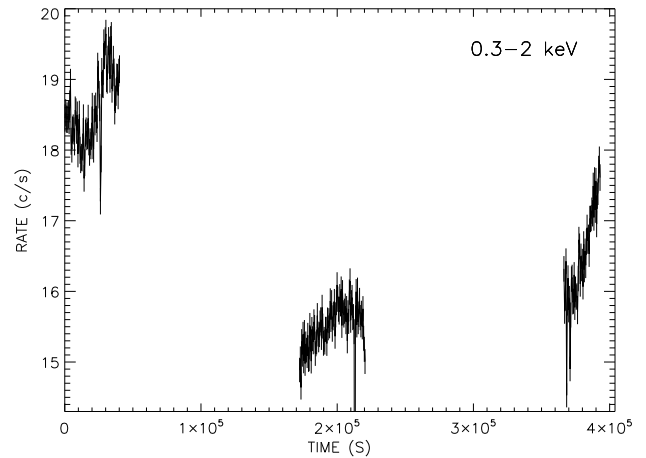}
\caption[]{EPIC-pn lightcurve in the 0.3-2~keV band.}
\label{lightcurve}
\end{figure}

Mrk 279 was observed by \XMM{} in three consecutive orbits between November 15th and November 19th 2005 for a total of $\sim$160~ks. The RGS data were processed using the standard pipeline {\tt SAS v9.0} (Gabriel et al. \cite{Gab04}). After filtering background flaring events, the RGS total net exposure time is $\sim$110~ks. A summary of the observation log is reported in Table~\ref{obslog}, and the lightcurve of the three observations is shown in Fig.~\ref{lightcurve}.

\begin{table}
  \centering
  \caption[]{\XMM{} observation log.}
  \label{obslog}

  \begin{tabular}{lcccc}
    \hline
    \hline
    \noalign{\smallskip}
    Orbit & Date & Net exp. time & RGS1 rate$^a$ & RGS2 rate$^a$ \\
          & (dd-mm-yy) & (ks) & (c/s) & (c/s) \\
    \noalign{\smallskip}
    \hline
    \noalign{\smallskip}
    1087 & 15-11-05 & 40.6 & 0.73 & 0.82 \\
    1088 & 17-11-05 & 42.5 & 0.61 & 0.68 \\ 
    1089 & 19-11-05 & 25.7 & 0.65 & 0.72 \\
    \noalign{\smallskip}
    \hline
    \noalign{\smallskip}
    \multicolumn{4}{l}{$^a$ 0.3-2.4~keV}\\
  \end{tabular}
\end{table}

It can be seen from Fig.~\ref{lightcurve} that the source experienced strong variations in flux in a relatively short time scale, which is especially clear in the plotted 0.3-2~keV band where we find a maximal variation in flux of about 20\% between orbits 1087 and 1088. The source flux recovered during orbit 1089 so that its brightness was comparable to that of orbit 1087 by the end of the observation. For clarity, in what follows the 1087-, 1088-, and 1089-orbit RGS datasets will be labeled as High ($H$), Low ($L$), and Recovering ($R$), respectively, attending to their relative flux state.

For the purpose of this study, the $R$ dataset was included in the overall fit described in Sect.~\ref{analysis} in order to increase the statistics. However, since we are interested in investigating the possible spectral variations in two clearly distinct flux states we will focus mainly on the $H$ and $L$ datasets. In all datasets under study, RGS1 and RGS2 were analysed together. We used C-statistics during the spectral analysis (see Sect.~\ref{analysis}) and therefore binning of the data is in theory no longer needed. However, to avoid oversampling, we rebinned the data by a factor of 3. The signal-to-noise ratio for both datasets is around $\sim$9 and $\sim$8 for $H$ and $L$, respectively, for the part of the wavelength region in which the most relevant spectral features fall.

%
\begin{figure}
\centering
\hbox{
\includegraphics[width=6.5cm,angle=90.]{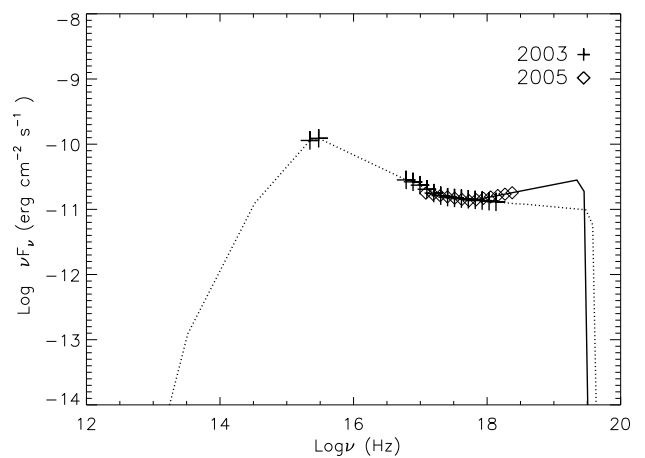}
}
\caption[]{Average spectral energy distribution of Mrk 279. The X-ray continuum is determined by \Chan{} (crosses, Costantini et al. \cite{cos07}). For comparison, we overplot the average continuum measured by \XMM{} (diamonds). For the far UV we relied on the {\it HST} and {\it FUSE} measurements.}
\label{sed}
\end{figure}


\begin{table}
  \centering
  \caption[]{Mrk 279 RGS continuum best-fit parameters.}
  \label{cont}

  \begin{tabular}{lcccc}
    \hline
    \hline
    \noalign{\smallskip}
    Dataset & $\Gamma$ & $F_{PL}$$^a$ & $T_{mbb}$ (keV)& $F_{mbb}$$^a$  \\
    \noalign{\smallskip}
    \hline
    \noalign{\smallskip}
    $H$ & 2.03$\pm$0.07 & 2.05$\pm$0.06 & 0.14$\pm$0.01 & 0.16$\pm$0.02 \\
    \noalign{\smallskip}
    \hline
    \noalign{\smallskip}
    $L$ & 1.89$\pm$0.08 & 1.72$\pm$0.05 & 0.14$\pm$0.01 & 0.16$\pm$0.02 \\ 
    \noalign{\smallskip}
    \hline
    \noalign{\smallskip}
    $R$ & 2.01$\pm$0.08 & 1.85$\pm$0.06 & 0.13$\pm$0.01 & 0.13$\pm$0.03 \\ 
    \noalign{\smallskip}
    \hline
    \noalign{\smallskip}
    \multicolumn{4}{l}{$^a$ 0.5-2~keV flux, in units of $10^{-11}$~erg cm$^{-2}$~s$^{-1}$}\\
  \end{tabular}
\end{table}

%

\begin{figure*}
\centering
\hbox{
\includegraphics[height=17cm,width=6.5cm,angle=90.]{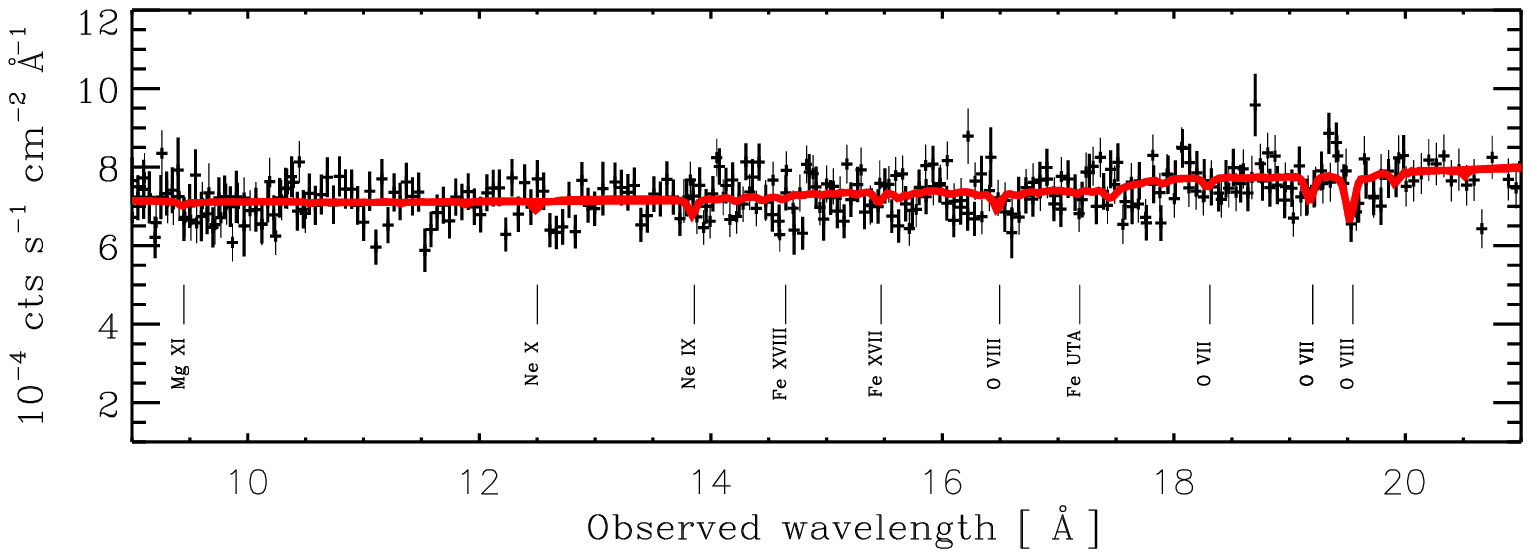}
}
\hbox{
\includegraphics[height=17cm,width=6.5cm,angle=90.]{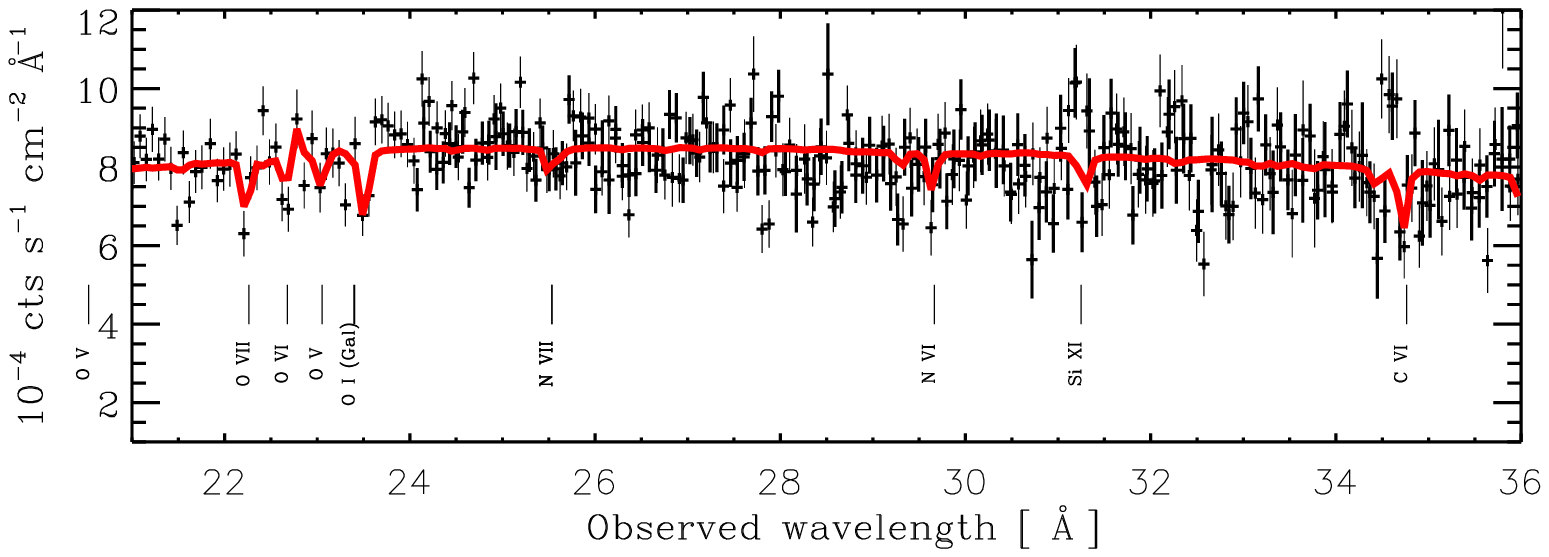}
}
\caption[]{RGS spectrum of Mrk 279. For plotting purposes all three observations have been stacked into a single spectrum. The solid line corresponds to the best-fit model. The most relevant spectral features are labeled.}
\label{spec_hilo}
\end{figure*}

\section{Spectral analysis with RGS}
\label{analysis}

The spectral analysis of the RGS data was carried out using the fitting package {\tt SPEX}\footnote{http://www.sron.nl/spex/} v2.0 (Kaastra et al. \cite{Kaa96}). Both RGS1 and RGS2 were fitted together, and C-statistics was the fitting method adopted (Cash \cite{Cash79}). The same spectral model was applied to $H$, $L$ and $R$ datasets and they were fitted simultaneously using three different sectors in {\tt SPEX}. The advantage of assigning a sector to each dataset under study resides in the fact that, although some of the parameters might be different in general (e.g. the continuum parameters), it allows parameters from different datasets to be coupled and fitted simultaneously (see {\tt SPEX} documentation for further information). The final C-stat/d.o.f.=4857/4381$\sim$1.1 obtained applies to the final simultaneous fit of all three datasets. In what follows we describe the continuum, and the absorption and emission features that are present in the spectrum of Mrk 279.

\begin{table*}
  \renewcommand{\arraystretch}{1.6}
  \centering
  \caption[]{Best-fit {\tt xabs} parameters to the warm absorber of Mrk 279.}
  \label{WAbest}

  \begin{tabular}{l|ccc|ccc|c}
    \hline
    \hline
    &  \multicolumn{2}{r}{\tt xabs1} & & \multicolumn{2}{r}{\tt xabs2} & \\
    \hline
     & $\log \xi$$^a$ & $N_{\rm{H}}$$^b$ & $v_{\rm{out}}$$^c$ & $\log \xi$$^a$ & $N_{\rm{H}}$$^b$ & $v_{\rm{out}}$$^c$ & C-stat/d.o.f. \\
    \hline
    $H$ dataset & 0.6$\pm$0.3 & 0.8$\pm$0.5 & -100$\pm$180 & 2.6$\pm$0.1 & 4.1$^{+2.7}_{-1.7}$ & -340$\pm$180 & 1642/1453 \\
    \hline
    $L$ dataset & 1.1$\pm$0.4 & 0.8$^{+0.8}_{-0.4}$ & -240$^{+145}_{-530}$ & 2.6$^{+0.2}_{-1.0}$ & 1.2$^{+1.7}_{-1.1}$ & -435$^{+575}_{-225}$ & 1539/1456 \\ 
    \hline
    $R$ dataset & 0.4$\pm$0.7 & 1.5$\pm$0.5 & -320$\pm$200 & 2.6$\pm$0.1 & 7.2$\pm$2.3 & -550$^{+285}_{-105}$ & 1681/1464 \\ 
    \hline
    Coupled datasets & 0.7$\pm$0.3 & 0.7$\pm$0.3 & -200$^{+90}_{-270}$ & 2.6$\pm$0.1 & 2.7$\pm$1.1 & -370$^{+90}_{-130}$ & 4857/4381 \\ 
    \hline
    \hline
    \noalign{\smallskip}
    \multicolumn{7}{l}{$^a$ In units of erg s$^{-1}$ cm$^{-2}$; $^b$ In units of $10^{20}$~cm$^{-2}$; $^c$ In units of km s$^{-1}$}\\
  \end{tabular}
\end{table*}

\subsection{Continuum spectrum}
\label{continuum}

The broadband (0.3-10~keV) continuum spectrum of Mrk 279 is best described by a broken power law and a black body (see Paper I). In the soft X-rays region (0.3-2~keV), however, a single power law provides a good description of the baseline continuum, although there exist deviations at very soft X-rays (soft excess).

To account for the soft excess we added a modified black body model ({\tt mbb} model in {\tt SPEX}) to the power law, which takes into account modifications of a simple black body by coherent Compton scattering and is based on the calculations of Kaastra \& Barr (\cite{KB89}). The improvement of the fit after adding this component was $\Delta$C = 72 for 6 degrees of freedom.

The best-fit parameters of the continuum are reported in Table~\ref{cont} for the $H$, $L$ and $R$ datasets. The slope $\Gamma$ seems slightly steeper when the source is in the high-flux state, although this value is consistent within the error bars with the value determined for the low-flux state. The value obtained for the $R$ dataset lies inbetween the former and it is still consistent within the error bars. The thermal component remains virtually unchanged in all three datasets.

\subsection{Absorption at redshift zero}
\label{galabs}

The continuum emission of Mrk 279 is furrowed by several absorption features, some of them consistent with absorption at redshift zero, revealing a complex absorption spectrum. The local absorbing gas seems to have two distinct components: a neutral or mildly ionised absorber, characterised by the \OI{} feature at $\sim$23.5~\AA, with a column density of $N_{\rm{H}}=1.64 \times 10^{20}$~cm$^{-2}$ (Elvis et al. \cite{Elvis89}), and an ionised component with $N_{\rm{H}}=3.6\pm0.3 \times 10^{19}$~cm$^{-2}$ and $kT=7.2\pm1.2$~eV (Costantini et al. \cite{cos07}).

We have modeled these components by means of a collisionally ionised plasma ({\tt hot} model in {\tt SPEX}) fixing their parameters to the values listed above. The ionised component likely originates in the interstellar medium of our Galaxy, but an extragalactic origin cannot be ruled out (Nicastro et al. \cite{Nic02}). Further discussion on the absorbing gas along the line of sight of Mrk 279 can be found in Williams et al. (\cite{Wil06}).

\subsection{The warm absorber}
\label{WA}

The spectrum of Mrk 279 shows a number of absorption features in the range $\sim$10-35~\AA~consistent with the presence of outflowing ionised gas at the redshift of the source. Previous observations of Mrk 279 with \Chan{} already showed that the absorbing gas consists of at least two phases with distinct ionisation degrees (Costantini et al. \cite{cos07}).

We have therefore modeled the absorption spectrum using two {\tt xabs} components in {\tt SPEX}. This model calculates the transmission of a slab of material where all ionic column densities are linked through a photoionisation balance model, which was calculated with {\tt CLOUDY} (Ferland et al. \cite{Fer98}) version C08. The input spectral energy distribution (SED) of Mrk 279 provided to {\tt CLOUDY} was based on the \Chan{} observation carried out in 2003, as the values obtained in the X-ray domain by both \Chan{} and the current \XMM{} observation are virtually the same (see Fig.~\ref{sed}). The mean velocity dispersion of the absorption lines was fixed to 50~km s$^{-1}$ (Costantini et al. \cite{cos07}).

\begin{figure}
\centering
\hbox{
\includegraphics[width=6.5cm,angle=90.]{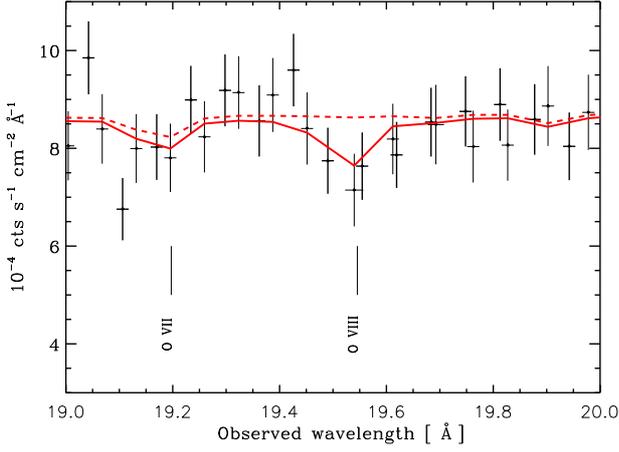}
}
\caption[]{Detail of the \OVIII{} absorption line. Overplotted is the fit with only one {\tt xabs} component (dashed line) and two {\tt xabs} components (solid line).}
\label{oviiizoom}
\end{figure}

The addition of one {\tt xabs} component improves the fit by $\Delta$C = 78 for 3 degrees of freedom but fails to reproduce the \OVIII{} Ly$\alpha$ line. A second {\tt xabs} component is therefore required, improving the fit by a further $\Delta$C/$\Delta \nu$=59/3. This model consisting of two {\tt xabs} components with distinct ionisation states provides a good fit to the observed data (see Fig.~\ref{oviiizoom}).

\begin{figure}
\centering
\hbox{
\includegraphics[width=6.5cm,angle=90.]{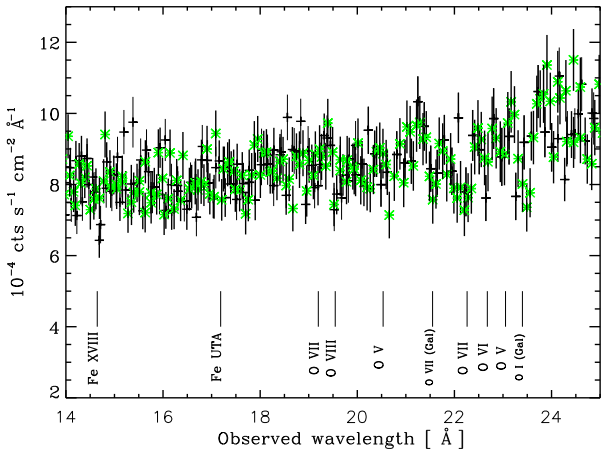}
}
\caption[]{Comparison between the $H$ (black crosses) and $L$ (light stars) datasets. The $L$ dataset is normalised to the unabsorbed continuum of the $H$ dataset.}
\label{ratio}
\end{figure}

We first fitted the warm absorber in the $H$, $L$ and $R$ datasets independently. If the variations in flux observed between the first two observations were propagated to the surrounding material, the ionisation properties of the gas should have changed accordingly. The results of the fits, however, show that no significant changes took place in the properties of the WA between the $H$ and $L$ flux states, as most of the parameters are consistent with each other within the error bars (see Table~\ref{WAbest}). Furthermore, we have normalised the $L$ dataset to the unabsorbed continuum of the $H$ dataset. In this way we get rid of any possible spectral effects caused by the change in flux between both observations so that only the information on the absorption features remains (see e.g. Netzer et al. \cite{Net03}). We can see in Fig.~\ref{ratio} that the oxygen complex as well as the most relevant iron absorption features remain virtually unchanged. The physical implications of this lack of response of the WA to the changes in the ionising flux between two consecutive observations will be further discussed in Sect.~\ref{discussion}. We must note, however, that although the variation in the average flux is of the order of $\sim$20\%, the uncertainties in $\xi$ are of the order of $\sim$50\% in the case of the lowest ionisation component. We cannot therefore discard the possibility that short scale variability did actually occur but is statistically undetected.

The results of the fits on the individual observations are reported in Table~\ref{WAbest}. The WA is clearly present, but its very shallow column density ($\sim 8\times 10^{19}$~cm$^{-2}$ and $\sim 4 \times 10^{20}$~cm$^{-2}$ for low and high ionised WA phases, respectively, in agreement with previous measurements) makes it rather difficult to detect. This is especially problematic in the $L$ dataset, where the low flux level make some of the absorption troughs to be almost undetectable, which is reflected in the larger error bars with respect to the $H$ dataset. The ionisation parameter $\xi$ of the high-ionisation phase seems to be surprisingly stable throughout the observations, seemingly insensitive to the flux variations of the source, although its value becomes rather uncertain in the $L$ dataset. Both WA phases seem to be outflowing at different velocities ($\sim200$~km~s$^{-1}$ and $\sim400$~km~s$^{-1}$), with the high-ionisation phase having a faster outflow. The fit to the $R$ dataset alone was severely affected by the low statistics due to the short exposure time, and it provided results consistent with those obtained for the $H$ and $L$ datasets, albeit with much larger error bars.
 
Since the WA parameters in all datasets were consistent within the error bars, we assumed that the WA was formally under the same physical conditions in both the high and low flux states. Therefore, we fitted again all three datasets coupling this time the $H$, $L$ and $R$ {\tt xabs} parameters to each other in order to reduce the associated error on the parameters under consideration. The result of this fit is also reported in Table~\ref{WAbest} and it will be the one we will use throughout this paper as a description of the WA in further calculations. This fit is also shown in Fig.~\ref{spec_hilo} where, for plotting purposes, we have stacked the spectra of all three observations. The spectra displayed in Fig.~\ref{spec_hilo} and the following figures were produced by dividing the photon spectra by the effective area of the detector so that no model-dependent fluxing was applied.

\subsection{Emission lines}
\label{EM}

In addition to the absorption features present in the soft X-ray spectrum of Mrk 279 there are also hints of emission features. Costantini et al. (\cite{cos07}) reported several narrow emission features of different elements and radiative recombination continua of \CV{}~and \CVI{}~found in the \Chan{}-LETG spectrum of Mrk 279. We were unable to detect most of these lines in the RGS datasets. We find marginal detection of the \OVIII{} Ly$\alpha$ and \OVII{} (forbidden) emission lines in the $L$ dataset, whereas in the $H$ dataset the higher continuum level swamps out these low-significance emission lines (see Table~\ref{emlintab}).

At the position of the \OVII{} triplet there is a weak broad emission line intrinsic to the source. The presence of this broad line was taken into account while fitting the spectrum of Mrk 279 by adding a Gaussian line to the model. The line is only marginally detected at the 90\% confidence level in the $H$ and $R$ datasets, whereas it is not significantly detected at all in the $L$ dataset (see Paper I for a detailed discussion). The complexities in this spectral range due to bad pixels and the \OV{}, \OVI{} and \OVII{} absorption lines are such that the significant detection of the \OVII{} broad line is difficult, considering the signal-to-noise ratio of the spectra. The origin of this line is likely the BLR and its luminosity can be accounted for by the LOC model (Baldwin et al. \cite{Bal95}). Further discussion on this broad feature and how it could be linked to the complex Fe K$\alpha$ region of Mrk 279 is given in Paper I.

We used the \OVII{}~forbidden emission line and the upper limits of the other emission lines, which provide mild constraints, to calculate the emission spectrum of the two-component WA (see Sect.~\ref{WA}) and potentially link the observed emission and absorption. In order to do this, we used {\tt CLOUDY} assuming a gas with a density of $10^8$~cm$^{-3}$. In this way, the slab of material remains geometrically thin relative to its distance from the central source. For the purpose of this calculation we used a covering factor of $f_c=1$, defined as the fraction of $4\pi$~sr covered by the gas as viewed from the central ionising source, i.e. the solid angle subtended by the gas. We note that this is different from the global covering factor, which is the fraction of emission intercepted by the absorber averaged over all lines of sight and it has been estimated to be about 0.5 (Crenshaw et al. \cite{Cren03}).

\begin{figure}
\centering
\hbox{
\includegraphics[width=6.5cm,angle=90.]{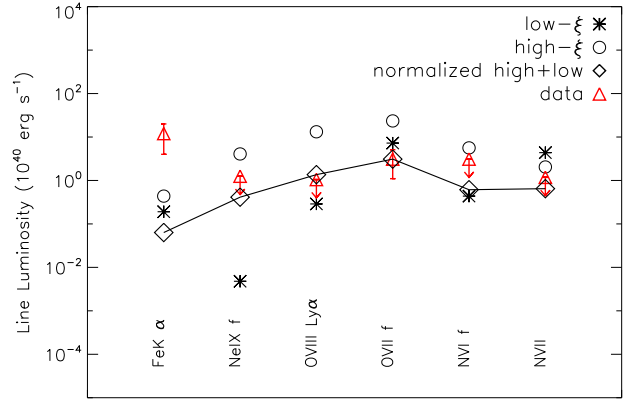}
}
\caption[]{Emission line contribution of the warm absorbers in Mrk 279, assuming a covering fraction $f_c=1$. Asterisks represent the low-ionisation component, circles the high-ionisation component, and triangles are the data. Connected diamonds represent the summed contribution scaled to the \OVII{} forbidden line luminosity ($f_c=0.1$, see text).}
\label{wa_low}
\end{figure}

In Fig.~\ref{wa_low} we display our results. Both gas components (asterisks and circles for the low- and high- ionisation components, respectively) tend to overestimate the measured luminosities (triangles) for $f_c=1$. This is especially clear for the high-$\xi$ component while it is less significant in the case of the low-$\xi$ component. However, the line luminosity linearly scales with $f_c$ in practice (this is not true for high column densities, when the dependence on $f_c$ is more complicated (Netzer \cite{Net93}, \cite{Net96}), but it is valid for shallow $N_H$ such as the ones found in this source). As the \OVII{} flux predicted by the high-ionisation component is approximately three times the flux of the low-ionisation gas, the scaling of the predicted-to-observed \OVII{} luminosity is dominated by the high-ionisation gas. If a single component were to explain the \OVII{} forbidden line, a $f_c$ of 0.4 and 0.13 would be needed for the low- and high-ionisation gas, respectively, whereas for a combination of both components a $f_c=0.1$ is required to match the luminosities predicted by {\tt CLOUDY} with the observed ones. This is in tune with what is found for unobscured or Compton-thin AGN, where absorption lines are dominant with respect to the emission lines in cases in which a cloud, belonging to a population of clouds with a low global covering factor, happens to lie along the line of sight of the observer. In this case the emission lines coming from other directions provide a negligible contribution to the total spectrum (Nicastro et al. \cite{Nic99}, Bianchi et al. \cite{Bianchi05}).

In Fig.~\ref{wa_low} we also plot the sum of the components scaled to the luminosity of the \OVII{} line (connected diamonds). We see that the normalised model is in agreement with the upper limits of the lower ionisation lines. We also note that none of the gas components that produce absorption are sufficient to produce the Fe K$\alpha$ narrow component in emission that is detected in the pn spectrum (see Paper I), being the total contribution of the order of $\sim$0.5\%.

\begin{table}
  \centering
  \caption[]{Column densitity and outflow velocity of the most relevant X-ray absorption lines.}
  \label{abslines}

  \begin{tabular}{lcc}
    \hline
    \hline
    \noalign{\smallskip}
    Ion & $\log N_{\rm{ion}}$  & $v_{\rm{out}}$  \\
       & (cm$^{-2}$)  & (km s$^{-1}$) \\
    \noalign{\smallskip}
    \hline
    \noalign{\smallskip}
    \CVI{}     &  17.16$\pm$0.11 & -420$\pm$60 \\
    \noalign{\smallskip}
    \NVI{}     &  16.68$\pm$0.31 & -340$\pm$110 \\ 
    \noalign{\smallskip}
    \NVII{}    &  16.24$\pm$0.32 & -450$^{+246}_{-219}$ \\
    \noalign{\smallskip}
    \OV{}      &  17.10$\pm$0.05 & -160$\pm$140 \\
    \noalign{\smallskip}
    \OVI{}     &  17.00$\pm$0.07 & -410$\pm$130 \\
    \noalign{\smallskip}
    \OVII{}    &  17.23$\pm$0.07 & -510$^{+140}_{-260}$ \\
    \noalign{\smallskip}
    \OVIII{}   &  17.40$\pm$0.13 & -70$^{+550}_{-100}$ \\
    \noalign{\smallskip}
    \NeIX{}    &  $<$16.2              & $\dots$ \\
    \noalign{\smallskip}
    \NeX{}     &  $<$16.6                 & $\dots$ \\
    \noalign{\smallskip}
    \SiXI{}    &  16.53$^{+0.70}_{-0.27}$ & -990$\pm$190 \\
    \noalign{\smallskip}
    \FeXVII{}  &  16.31$\pm$0.16 & -265$^{+220}_{-190}$ \\
    \noalign{\smallskip}
    \FeXVIII{} &  15.86$\pm$0.16 & -1250$\pm$315 \\
    \noalign{\smallskip}
    \hline
  \end{tabular}
\end{table}

\begin{table}
  \centering
  \caption[]{Emission lines luminosities measured by RGS and their significance.}
  \label{emlintab}

  \begin{tabular}{lcccc}
    \hline
    \hline
    \noalign{\smallskip}
    Ion & Rest wavelength & $L_{obs}$ & F-test & Model \\
        & (\AA) & ($10^{40}$~erg s$^{-1}$) & (\%) & \\
    \noalign{\smallskip}
    \hline
    \OVIII{} Ly$\alpha$ & 18.97 & $<$1.0 & $\dots$ & {\tt delta}\\
    \OVII{} (f) & 22.10 & 3.1$\pm$2.0 & 89.0 & {\tt delta}\\ 
    \OVII{} (triplet) & 21.9 & 13$\pm$7 & 95.0 & {\tt gauss}  \\
    \hline
  \end{tabular}
\end{table}


\section{Discussion}
\label{discussion}

\subsection{Two ionisation components in the warm absorber}
\label{WAion}

The absorption features detected in the spectrum of Mrk 279 are consistent with those produced by different elements, mainly oxygen, carbon, nitrogen and iron, with different degrees of ionisation. These troughs are blueshifted with respect to the systemic velocity of the source, and their degree of ionisation reveals that they are associated with a warm absorber often seen in other Seyfert 1 galaxies. The absorbing gas presents at least two phases with ionisation parameters of $\log \xi \sim 0.7$ and $\log \xi \sim 2.6$, respectively, and quite low $N_H$ (of the order of $\sim 10^{20}$~cm$^{-2}$, see Table~\ref{WAbest}).

\begin{figure}
\centering
\hbox{
\includegraphics[width=6.5cm,angle=-90.]{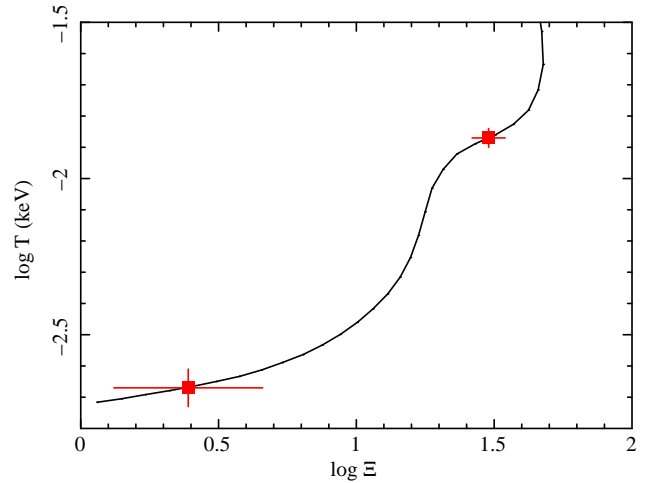}
}
\caption[]{Pressure ionisation parameter as a function of electron temperature. The values corresponding to the warm absorber components in Mrk 279 are marked as squares.}
\label{xitemp}
\end{figure}

The question about whether the outflowing gas forms clouds in equilibrium with its surroundings or it is part of a continuous distribution of ionised material is difficult to answer in the case of Mrk 279. In the first case, the clouds should be in pressure equilibrium with the less dense environment in order to grant long-lived structures. Such a scenario has been already claimed in the case of the Seyfert 1 galaxy NGC 3783 (Krongold et al. \cite{Kron03}, Netzer et al. \cite{Net03}, Krongold et al. \cite{Kron05}). If the WA components belong indeed to a discrete scenario with a sharp bimodal distribution then their pressure ionisation parameter $\Xi$ should be the same. $\Xi$ can be defined as a function of pressure or temperature: $\Xi = L/4\pi r^2cp=\xi/4\pi ckT$. If we overplot the values of $\Xi$ corresponding to the two phases of the WA detected in Mrk 279 on the $\Xi - T$ curve of the source we can see that both components cannot be in pressure equilibrium unless other processes such as magnetic confinement are playing a role (see Fig.~\ref{xitemp}). Similar results were also found in the \Chan{}~analysis of Mrk 279 (Costantini et al. \cite{cos07}).


\begin{figure*}
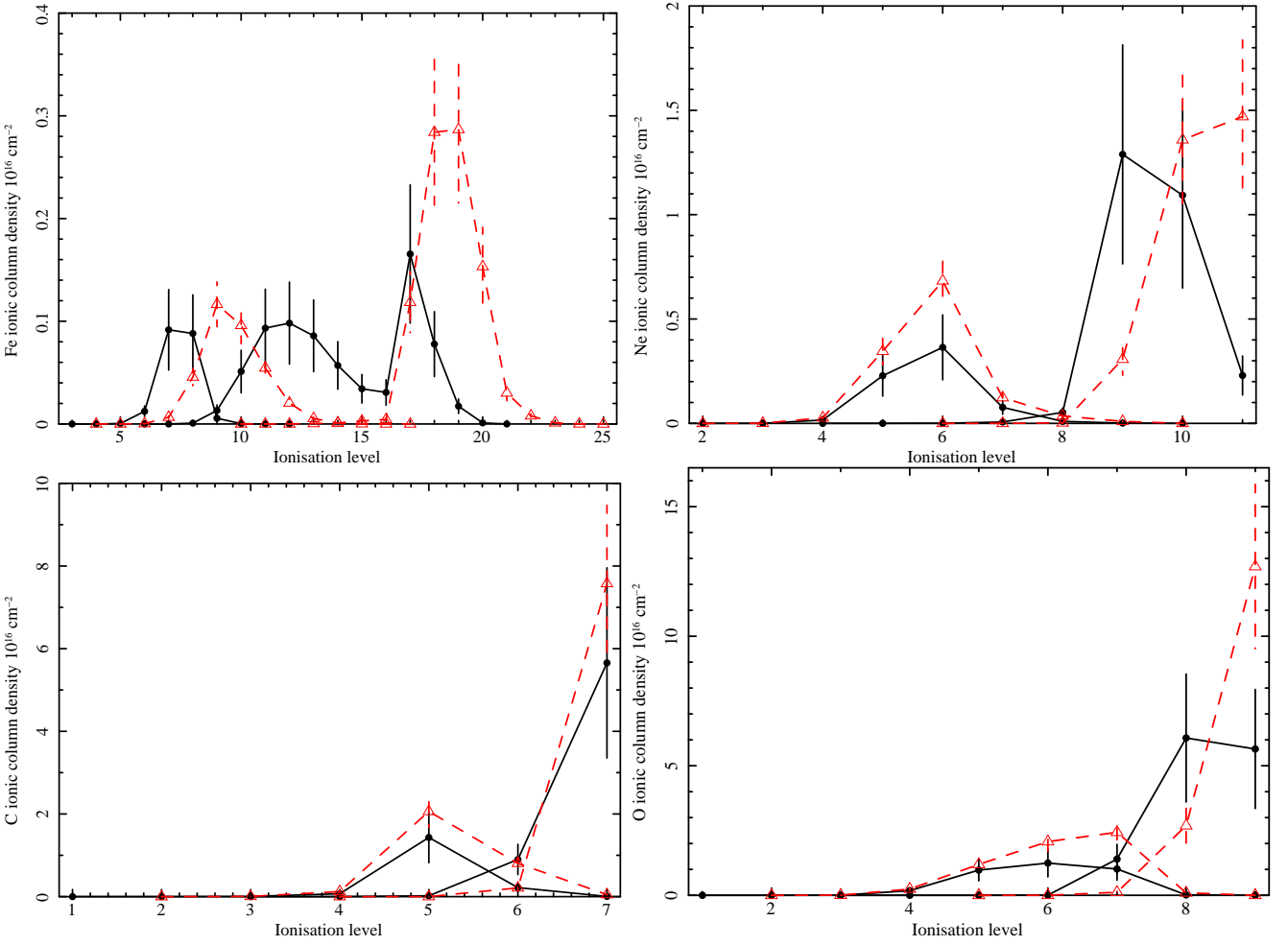

\centering
\hbox{
\includegraphics[width=6.5cm,angle=-90.]{Fe_col_new.ps}
\includegraphics[width=6.5cm,angle=-90.]{Ne_col_new.ps}
}
\hbox{
\includegraphics[width=6.5cm,angle=-90.]{C_col_new.ps}
\includegraphics[width=6.5cm,angle=-90.]{O_col_new.ps}
}
\caption[]{Ionic column densities of Fe ({\it upper left panel}), Ne ({\it upper right panel}), C ({\it lower left panel}), and O ({\it lower right panel}) ions in the \XMM{} (solid lines, this work; dots: low-$\xi$ ions, triangles: high-$\xi$ ions) and \Chan{} (dashed lines, Costantini et al. \cite{cos07}; squares: low-$\xi$ ions, crosses: high-$\xi$ ions) observations.}
\label{ioncols}
\end{figure*}

\subsection{Is the warm absorber variable?}
\label{WAdens}

Mrk 279 is known for experiencing moderate flux variability on timescales of a few tens of ks (Scott et al. \cite{Scott04}, Costantini et al. \cite{cos09}). If these changes in the ionising flux were to produce any measurable changes in the physical conditions of the absorbing gas on both short and long timescales, they could in principle be used to estimate the location of the absorber.

The \XMM{} lightcurve of Mrk 279 in the 0.3-2~keV range looked promising in this respect, since we observe a flux variation of $\sim$20\% between the $H$ and $L$ observations in the soft X-ray band (Fig.~\ref{lightcurve}). If the WA were in equilibrium the ionisation parameter would then vary accordingly in response to the changes in the continuum. Unfortunately, given the quality of the data and the shallow column density of the WA such changes are difficult to detect (if present at all). Since no differences in the properties of the WA were significantly measured (see Sect.~\ref{WA}) this could mean that either the WA was not in equilibrium yet, i.e. the gas did not have enough time to recombine after the change in the ionising flux and did not reach the equilibrium again, or that the changes actually took place but we were unable to detect them given the present statistics.

Alternatively there is the possibility of observing changes in the WA at longer timescales comparing observations several years apart, which would enable us to put an upper limit to the location of the absorber. For this purpose we compared the current \XMM{} observation with the one by \Chan{} two and a half years earlier. However, the flux state of Mrk 279 was not strikingly different between both observations (Costantini et al. \cite{cos07}), and the SED in both observations is virtually identical in the X-ray domain (Fig.~\ref{sed}). In these circumstances the grounds for a variability study are rather weak.

A noticeable difference was the detection of radiative recombination continua (RRC) of \CV{}~($\lambda_{obs}=32.57$~\AA) and \CVI{}~($\lambda_{obs}=26.05$~\AA), which are clear signatures of recombination in overionised plasmas, in the \Chan{} spectrum of Mrk 279 (Costantini et al. \cite{cos07}), whereas no hints of such features were found in the \XMM{} spectra. We tried nevertheless to obtain upper limits to the emission measure of these features by fixing their temperature to the values found in Costantini et al. (\cite{cos07}), $kT = 1.7$~eV. We found $EM < 0.25 \times 10^{61}$~cm$^{-3}$ and $EM < 0.40 \times 10^{61}$~cm$^{-3}$ for the \CV{} and \CVI{} RRCs, respectively. These values are lower than those reported in Costantini et al. (\cite{cos07}), which suggests that the present spectrum is the result of a different flux history with respect to that of \Chan{}. Indeed Mrk 279 is known to significantly vary (up to a factor of 2) on both short and long timescales.

We looked for possible changes in the WA by looking at the ionic column densities measured by \Chan{} and \XMM{}. We obtained the column densities of several ions from the {\tt xabs} components used in our model. In Figure~\ref{ioncols} we plot the ionic column densities obtained from the best fit to the WA of this work (solid lines) and those obtained by Costantini et al. (\cite{cos07}) with \Chan{} (dashed lines) for different elements whose absorption imprints are present in the spectrum of Mrk 279. The errors on the column density of each ion are rescaled with respect to that of $N_H$. Only in the case of Fe and Ne we can clearly trace both components of the WA, seen as peaks in the ionic column densities of these elements.

In order to statistically assess whether the ionic column densities varied significantly between the \Chan{} and \XMM{} observations we applied a Kolmogorov-Smirnoff test on both $N_{\rm{ion}}$ distributions. We did not find significant changes in the ionic column densities of Ne, Fe, C and O. The significancies of the variations were only at the 90\% confidence level or less ($P_{\rm{KS}}=81$\% for carbon, $P_{\rm{KS}}=90.6$\% for oxygen, $P_{\rm{KS}}=88$\% for iron, and $P_{\rm{KS}}=91$\% for neon). Therefore, we cannot affirm that the WA in Mrk 279 experienced either long- or short-term variations in response to the changes in the ionising continuum.

%
\begin{figure}
\centering
\hbox{
\includegraphics[width=6.cm,angle=-90.]{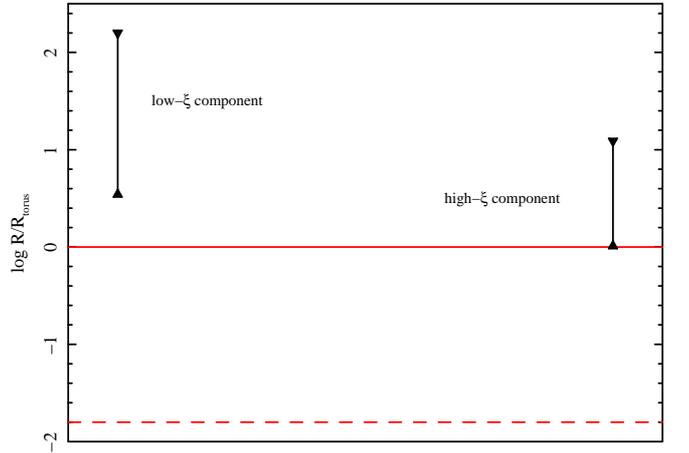}
}
\caption[]{Estimated minimum and maximum distances of both WA phases in Mrk 279 from the central engine in units of the torus distance $R_{torus}$. The solid line represents $R=R_{torus}$, while the dashed line represents the ratio of the BLR distance $R_{BLR}$ to the torus distance.}
\label{Rplot}
\end{figure}

\subsection{The location of the warm absorber}
\label{WAloc}

Even though we have not significantly measured any variations in the WA, which we could have used to break the degeneracy of the electron density $n_e$ and the distance $R$, we can still put mild constraints on the location of the absorbing gas. For this purpose we will follow the same argumentation given in Blustin et al. (\cite{Blustin05}, B05 hereafter) for a sample of 23 AGN.

The main assumption for estimating the maximum distance to the base of the WA from the central engine is that most of the mass is concentrated in a layer of depth $\Delta r$ so that it is less or equal to the distance to the central source $R$: $\frac{\Delta r}{R}\leq 1$. On the other hand, the column density observed along our line of sight is a function of the density of the gas, which is generally dependent on the distance $n(R)$, its volume filling factor $C_v$, and $\Delta r$: $N_H \sim n(R)C_V\Delta r$.

The ionisation parameter $\xi$ is defined by the ratio of the ionising flux and the density of the gas: $\xi=\frac{L}{n(R)R^2}$, where $L$ is the 1-1000 Rydberg luminosity, $n(R)$ is the density, and $R$ is the distance from the ionising source. Combining the equations above we obtain:

\begin{equation}
\label{deltar2}
\frac{\Delta r}{R}\sim \frac{\xi R N_H}{L_{ion}C_V},
\end{equation}

\noindent so that after applying the $\frac{\Delta r}{R}\leq 1$ condition:

\begin{equation}
\label{rmax}
R \leq \frac{L_{ion}C_V}{\xi N_H}.
\end{equation}

The volume filling factor $C_V$ was estimated assuming that the momentum of the outflow must be of the order of the momentum of the absorbed radiation plus the momentum of the scattered radiation (B05):

\begin{equation}
\label{momentum}
\dot{M}v \sim \dot{P}_{abs}+\dot{P}_{sc} = \frac{L_{abs}}{c}+\frac{L_{ion}}{c}(1-e^{-\tau_T}),
\end{equation}

\noindent where $\dot{M}$ is the mass outflow rate, $L_{abs}$ is the luminosity absorbed by the outflow in the 1-1000 Ryd range, $c$ is the speed of light, and $\tau_T=\sigma_TN_H$ is the optical depth for Thomson scattering.

The expression we used for the estimation of the mass outflow rate is the Eq. 18 in B05, which was deduced assuming that the outflowing mass is contained within a segment of a thin spherical shell and that it has cosmic elemental abundances ($\sim$75\% by mass of hydrogen and $\sim$25\% by mass of helium):

\begin{equation}
\label{cv}
\dot{M} \sim \frac{1.23m_pL_{ion}C_Vv\Omega}{\xi},
\end{equation}

\noindent where $m_p$ is the mass of the proton, $L_{ion}$ is the 1-1000 Rydberg ionising luminosity, $C_V$ is the volume filling factor, $v$ is the velocity of the outflow, $\Omega$ is the solid angle subtended by the outflow, and $\xi$ is the ionisation parameter of the outflow.

$L_{ion}$, $\xi$ and $v$ are directly measured from this observation. For each component of the WA we used the values reported in Table~\ref{WAbest} for the coupled datasets, and $L_{ion}=2.4 \times 10^{44}$~erg~s$^{-1}$. The opening angle of the outflow is difficult to estimate and we applied the average value of $\Omega \sim 1.6$ used in B05 assuming that 25\% of the nearby AGN are type 1 (Maiolino \& Riecke \cite{MR95}), and that the covering factor of these outflows seems to be $\sim$50\% (e.g. Reynolds \cite{Rey97}). Note that this value yields $\Omega/4\pi=0.13$, in extraordionary agreement with what we found in Sect.~\ref{EM}.

Combining Eqs.~\ref{cv} and~\ref{momentum} an expression for the volume filling factor is obtained\footnote{Note that in Eq. 23 in Blustin et al. (\cite{Blustin05}) there was a typo in the denominator of the equation: $c$ should not be there}:

\begin{equation}
\label{cv2}
C_V \sim \frac{(\dot{P}_{abs}+\dot{P}_{sc})\xi}{1.23m_pL_{ion}v^2\Omega}.
\end{equation}

\noindent Applying this equation to both components of the WA we find a volume filling factor of $C_V \sim 1.1\times10^{-3}$ and $C_V \sim 0.025$ for the low- and high-$\xi$ components, respectively. These low values may suggest that the WA is likely to consist of filaments or clouds passing through our line of sight. In the sample of 23 AGN of B05 none of the estimated volume filling factors were higher than $\sim$8\%.

Solving Eq.~\ref{rmax} for each of the WA components in Mrk 279 tells us that the high-ionisation phase can extend up to $\sim$20~pc from the central source, whereas the low-ionisation component can extend up to $\sim$260~pc.

A lower limit to $R$ is harder to reliably estimate in these circumstances. A very mild constraint can be obtained using dynamical arguments. If we assume that the measured outflow velocities should be greater than or equal to the escape velocity from the AGN we have:

\begin{equation}
\label{escvel}
v_{\rm{out}}\geq \sqrt{\frac{2GM}{R}},
\end{equation}

\noindent where $G$ is the constant of gravity, $M$ is the mass of the supermassive black hole and $R$ is the distance of the WA from the black hole. The mass of the central black hole in Mrk 279 is $M=2.57\times 10^7$~$M_\odot$ (Wandel \cite{Wan02}). This gives us a lower limit estimate of $R \gtrsim 1.6$~pc and $R \gtrsim 5.5$~pc for the high- and low-$\xi$ components, respectively. We note that assuming that the outflowing gas is escaping the gravitational well of the AGN is not a strong constraint indeed. The gas could be well under the escape velocity and either dissipate or become completely ionised before it starts inflowing again towards the central black hole.

With these caveats in mind, both WA components in Mrk 279 seem to be located further away than the BLR, which is at $r_{BLR}\sim 30$~days-light according to the relation given in Wandel (\cite{Wan02}), and it is likely associated with the putative dusty torus. Using the approximate relation of Krolik \& Kriss (\cite{KK01}) for the inner edge of the torus $r_{\rm{inner}}\sim 1 \times L^{0.5}_{44}$ in pc, where $L_{44}$ is the 1-1000 Rydberg luminosity in units of $10^{44}$~erg s$^{-1}$, we obtain $r_{\rm{inner}}\sim 1.6$~pc. In Fig.~\ref{Rplot} we plot these estimates relative to the torus distance and the BLR. Similar results were found for the compilation of 23 AGN of B05, where it can be seen that the minimum and maximum WA distances for the Seyferts and NLSy1s are consistently further out than the BLR and tend to cluster around the location of the torus. This would rule out an accretion disc wind origin for these absorbers and would be closer to the torus wind origin proposed by Krolik \& Kriss (\cite{KK01}).

\subsection{Energetics of the warm absorber}
\label{WAener}

It would be interesting to know whether the outflows commonly associated to the WA have an impact on the interestellar medium of the host galaxy of the AGN or even further out in the intergalactic medium. We can estimate how much mass is carried out of the innermost parts of the AGN by means of the outflows and how it compares with the accreted matter onto the central supermassive black hole.

From Eq.~\ref{cv} we find $\dot{M}\sim 0.057$~M$_\odot$~yr$^{-1}$ and $\dot{M}\sim 0.031$~M$_\odot$~yr$^{-1}$, for the low- and high-$\xi$ components, respectively. These values are consistent with those of the WA in other Seyfert 1 galaxies (see Blustin et al. \cite{Blustin05}, and references therein).

The mass accretion rate in Mrk 279 is calculated using:

\begin{equation}
\label{macc}
\dot{M}_{acc}=\frac{L_{bol}}{\eta c^2},
\end{equation}

\noindent where we assume a nominal accretion efficiency of $\eta=0.1$ and a bolometric luminosity of $L_{bol}=7.56\times10^{43}$~erg~s$^{-1}$ (Dasyra et al. \cite{Das08}). Hence, the mass accretion rate is $\dot{M}_{acc}=0.0133$~M$_\odot$~yr$^{-1}$. The outflow rate is higher than the accretion rate in the case of Mrk 279, which is commonly found in other type 1 AGN. For instance, 9 out of the 13 type 1 AGN in the B05 sample also present this behaviour.

From the outflow rate we can estimate how important are these outflows in energetic terms. The kinetic luminosity (kinetic energy carried out by the outflow per unit time) is defined as:

\begin{equation}
\label{lke}
L_{KE}=\frac{1}{2}\dot{M}_{out}v^2.
\end{equation}

\noindent Using the $\dot{M}_{out}$ values and outflow velocities found for both low- and high-$\xi$ components in our WA we obtain kinetic luminosities of $L_{KE}=7.25\times10^{38}$~erg~s$^{-1}$ and $L_{KE}=1.30\times10^{39}$~erg~s$^{-1}$, which account for $\sim$0.001\% and $\sim$0.002\% of the bolometric luminosity, respectively. Similarly, the kinetic luminosities estimated for the B05 sample represent in all the cases less than 1\% of the bolometric luminosity of the AGN. Therefore, these outflows play a limited role in the energetics of the system.

Given these $L_{KE}$ values, the kinetic energy injected to the medium would be $E_{KE} \sim L_{KE}t \sim 10^{55}$~erg, assuming that the average AGN lifetime is of the order of $t \sim 10^{8-9}$~years (Ebrero et al. \cite{Ebrero09}, Gilli et al. \cite{Gilli09}). For instance, the energy needed to disrupt the ISM of a typical galaxy is estimated to be of the order of $\sim 10^{57}$~erg (Krongold et al. \cite{Kron10}), which means that the outflows in Mrk 279 are not powerful enough to critically affect the ISM of the host galaxy.

\subsection{Abundances in the warm absorber}
\label{WAabun}

We calculated the abundances of various elements in the warm absorber of Mrk 279. The computed values were provided by {\tt xabs} in {\tt SPEX} and were calculated relative to a default set of standard abundances. In this work we have used the recent compilation of proto-solar abundances by Lodders ({\cite{Lod03}}). Our results are reported in Table~\ref{abun}. We find the element abundances to be consistent with the Solar ones in the cases of nitrogen and iron, carbon and oxygen.

Arav et al. (\cite{Arav07}) calculated the C, N and O abundances in the WA of Mrk 279 using deep and simultaneous observations of Mrk 279 by {\it FUSE} and {\it HST-STIS}. The calculations of Arav et al. (\cite{Arav07}) were done using {\tt CLOUDY} v96b5 and the standard set of abundances was the default of that version of {\tt CLOUDY} (Allende Prieto et al. \cite{AP01} for oxygen, Allende Prieto et al. \cite{AP02} for carbon, and Holweger \cite{Hol01} for nitrogen). The differences between this set of standard abundances and that of Lodders (\cite{Lod03}) is below 1\%, and therefore we can compare our abundance determinations using Lodders with the results of Arav et al. (\cite{Arav07}) finding that our results are fully consistent with theirs within the error bars with the exception of the nitrogen abundance, which was found to be slightly supersolar in Arav et al. (\cite{Arav07}). On the other hand, we cannot confirm the significant supersolar abundances reported by Fields et al. (\cite{Fields07}) using \Chan{} observations of Mrk 279.

\subsection{The lower ionisation emission lines}
\label{lowionem}

In the RGS spectrum, only the \OVII{} narrow forbidden line at 22.10~\AA~is detected. For other important He-like and H-like ions we only obtain upper limits (see Table~\ref{emlintab}). We find that emission from the WA gas can easily reproduce the \OVII{} line and, at the same time, be consistent with the upper limits on the other X-ray lines, provided that the covering factor of the gas is $\sim$0.1 (see Sect.~\ref{EM}). This result is not unexpected as in Type 1 objects, where the continuum dominates the emission, the absorption features of the ionised gas along our line of sight are always more evident than the emission lines. For instance, the \OVII{} forbidden line does not display any outflow velocity but, as the measured blueshift for the absorption line is only $\sim$370~km s$^{-1}$, this would not be evident in the emission component, which is directed in a wider angle.

Alternatively, \OVII{} might come from reflection from the distant torus itself. In order to test this hypothesis we used the model by Ross \& Fabian (\cite{RF05}) where reflection by constant-density and optically-thick material is considered. The calculation includes a number of ions, in addition to iron ({\tt REFLION} model in {\tt XSPEC}, Arnaud \cite{Arn96}). In particular, we used a modified version of {\tt REFLION} where values lower than $\xi=30$ can be used (G. Miniutti, private communication) in order to reproduce high \OVII{} to \OVIII{} ratios. Indeed gas with $\xi=30$ ($\log \xi=1.47$) already produces a considerable amount of \OVIII{}, which is not representative of our data. We note that the model cannot be tested directly on the RGS data, as the intrinsic resolution of the model grid is of only 0.3~\AA, which is not appropiate for high-resolution ($\Delta \lambda/\lambda \sim 0.05$) data. We then fitted the EPIC-pn spectrum of Mrk 279 using {\tt XSPEC} and found the best-fit parameters for the reflection component. We then evaluated the intrinsic line luminosities of \OVII{} and \OVIII{} and compared them with the real measurements. Although the iron line at 6.4~keV can be fitted by the reflection model, the predicted soft energy lines are not well explained. We found that, within the allowed range of best-fit $\xi$ values ($\xi < 27$), the soft energy line luminosities are always overpredicted by the model, by even a factor of 10 or more. This underlines the importance of \OVII{} and \OVIII{} as diagnostic lines for cold reflection modelling.


\begin{table}
  \centering
  \caption[]{Abundances in the warm absorber of Mrk 279.}
  \label{abun}

  \begin{tabular}{lc}
    \hline
    \hline
    \noalign{\smallskip}
    Element & Abundance$^a$ \\
    \noalign{\smallskip}
    \hline
    \noalign{\smallskip}
    Carbon   & 1.07$\pm$0.50   \\
    \noalign{\smallskip}
    Nitrogen & 1.07$^{+0.56}_{-0.48}$ \\ 
    \noalign{\smallskip}
    Oxygen   & 1.14$\pm$0.31    \\
    \noalign{\smallskip}
    Iron     & 1.11$\pm$0.44 \\
    \noalign{\smallskip}
    \hline
    \noalign{\smallskip}
    \multicolumn{2}{l}{$^a$ Relative to Solar, set to Lodders (\cite{Lod03})}\\
  \end{tabular}
\end{table}


\section{Conclusions}
\label{conclusions}

We have analysed the absorption features present in the \XMM{}-RGS spectrum of Mrk 279, consistent with the existence of a warm absorber at the redshift of the source. The absorbing material shows two phases with different degrees of ionisation, $\log \xi = 0.7\pm0.3$ and $\log \xi = 2.6\pm0.1$, and very shallow column densities ($N_{\rm{H}}=0.7\pm0.3\times 10^{20}$~cm$^{-2}$ and $N_{\rm{H}}=2.7\pm1.1 \times 10^{20}$~cm$^{-2}$, respectively).

The source was observed by \XMM{} in three consecutive orbits during which it experienced a significant drop in brightness (up to a 20\%) followed by a recovery. The WA did not show any significant variations in its parameters in response to the changes in flux. Likewise, no hints of significant long-term variability were found in Mrk 279 when comparing the current \XMM{} and past \Chan{} observations. The best-fit parameters of the WA in both observations were found to be consistent within the error bars and a K-S test on the distribution of ionic column densities of C, O, Ne and Fe between both observations only provided significancies lower than the 90\% confidence level.

Following the argumentation of Blustin et al. (\cite{Blustin05}) we were able to put mild constraints on the possible location of the WA, finding that both components are likely located between a few pc and up to $\sim$20~and $\sim$260~pc for the high- and low-ionisation phases, respectively. This would mean that the WA in Mrk 279 is likely to be associated to the putative dusty torus rather than to the BLR or winds arising from the accretion disc. The energetics of the WA outflows show that they have little or no impact in the sourrounding ISM of the Mrk 279 host galaxy. Abundances in the WA outflow were found to be consistent with Solar. The \OVII{} narrow forbidden line can be formed in the same gas that causes absorption at soft energies, provided that the covering factor of the gas, defined as the fraction of $4\pi$~sr covered by the absorber as seen from the central source, is about 0.1. This line is inconsistent with production by reflection in the molecular torus.

\begin{acknowledgements}
The Space Research Organization of The Netherlands is supported financially by NWO, the Netherlands Organization for Scientific Research. We thank the anonymous referee for comments that improved this paper.
\end{acknowledgements}

\end{document}